\documentclass[%
 aip,
 amsmath,amssymb,
 reprint,%
]{revtex4-1}

\usepackage[latin1]{inputenc}
\usepackage{rotating}
\usepackage{caption2}
\usepackage{bm}        
\usepackage{amssymb}   
\usepackage{amsmath}
\usepackage{threeparttable}
\usepackage{cancel}
\usepackage{adjustbox}
\usepackage{tablefootnote}
\usepackage[version=3]{mhchem}
\usepackage{tabularx}
\usepackage[T1]{fontenc}
\usepackage{adjustbox}
\usepackage{tablefootnote}
\usepackage{booktabs, makecell, multirow, threeparttable}
\usepackage{longtable}
\newcommand{\Lower}[1]{\smash{\lower 1.5ex \hbox{#1}}}   
\newcommand{\STAB}[1]{\begin{tabular}{@{}c@{}}#1\end{tabular}}

\newcommand*{\onet}{\textbf{1t}}
\newcommand*{\twot}{\textbf{2t}}
\newcommand*{\threet}{\textbf{3t}}
\newcommand*{\oneo}{\textbf{1o}}
\newcommand*{\twoo}{\textbf{2o}}
\newcommand*{\threeo}{\textbf{3o}}
\begin{document}

\title{The relationship between structure and excited-state properties in polyanilines from geminal-based methods.}

\author{Seyedehdelaram Jahani}
\affiliation{Institute of Physics, Faculty of Physics, Astronomy and Informatics, Nicolaus Copernicus University in Toru\'n, Grudziadzka 5, 87-100 Torun, Poland}

\author{Katharina Boguslawski}
\affiliation{Institute of Physics, Faculty of Physics, Astronomy and Informatics, Nicolaus Copernicus University in Toru\'n, Grudziadzka 5, 87-100 Torun, Poland}

\author{Pawe{\l} Tecmer}\email{ptecmer@fizyka.umk.pl}
\affiliation{Institute of Physics, Faculty of Physics, Astronomy and Informatics, Nicolaus Copernicus University in Toru\'n, Grudziadzka 5, 87-100 Torun, Poland}

\begin{abstract}
We employ state-of-the-art quantum chemistry methods to study the structure-to-property relationship in polyanilines (PANIs) of different lengths and oxidation states. 
Specifically, we focus on leucoemeraldine, emeraldine, and pernigraniline in their tetramer and octamer forms.
We scrutinize their structural properties, HOMO and LUMO energies, HOMO-LUMO gaps, and vibrational and electronic spectroscopy using various Density Functional Approximations (DFAs). 
Furthermore, the accuracy of DFAs is assessed by comparing them to experimental and wavefunction-based reference data. 
We perform large-scale orbital-optimized pair-Coupled Cluster Doubles (oo-pCCD) calculations for ground and electronically excited states and conventional Configuration Interaction Singles (CIS) calculations for electronically excited states in all investigated systems. 
The EOM-pCCD+S approach with pCCD-optimized orbitals allows us to unambiguously identify charge transfer and local transitions across the investigated PANI systems---an analysis not possible within a delocalized canonical molecular orbital basis obtained, for instance, by DFAs.
We show that the low-lying part of the emeraldine and pernigraniline spectrum is dominated by charge transfer excitations and that polymer elongation changes the character of the leading transitions. 
Furthermore, we augment our study with a quantum informational analysis of orbital correlations in various forms of PANIs.
\end{abstract}

\maketitle


\section{Introduction}
Organic-based semiconductors are essential building blocks of organic electronic devices, such as field-effect transistors, light-emitting diodes, memory cells, solar cells, and sensors.~\cite{salikhov2018new} 
The research progress in organic electronics has been greatly accelerated by the discovery of conducting polymers in 1977.~\cite{conductive-polymer-discovery}
The importance of this scientific discovery led to the 2000 Nobel prize in chemistry "for the discovery and development of conductive polymers".~\cite{shirakawa-nobel-lecture} 
Among the conducting polymers, the most studied are polyanilines (PANIs). 
Due to their environmental stability,~\cite{kulkarni1989thermal, kulkarni1991thermal} cost-effectiveness, ease of synthesis,~\cite{li2009polyaniline} and controllable electrical conductivity,~\cite{mishra2015dft, ray1989polyaniline} PANIs became a very popular conducting polymer. 
PANIs find applications in catalysis,~\cite{chen2012nanostructured, wu2012nitrogen} energy storage,~\cite{silakhori2013accelerated} battery electrode materials,~\cite{liu2013hierarchical} sensors,~\cite{ates2013review} and solar cells.~\cite{ameen2010sulfamic, zou2009photosynthetic} 
PANIs usually act as a donor and the fullerene containing-unit as an acceptor in the latter. 
Thus, the PANIs' Highest Occupied Molecular Orbital (HOMO) energy level dictates the electron-donating properties.

\begin{figure}
\centering
    \includegraphics[scale=0.23]{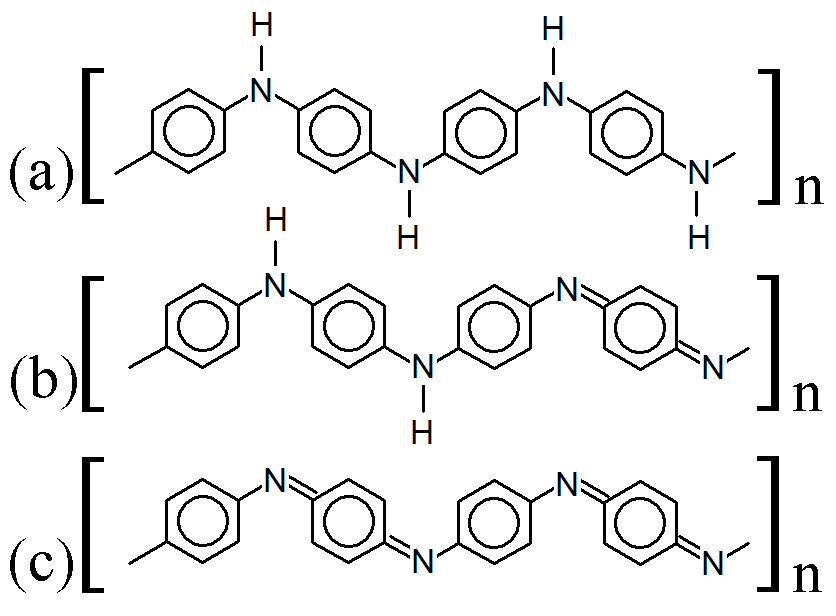}
\caption{Lewis structures of PANIs. (a) leucoemeraldine, (b) emeraldine, and (c) pernigraniline drawn with ChemPlot.~\cite{chemplot}}\label{fig:PANI-lewis-structures}
\end{figure}

What distinguishes PANIs from other conducting polymers is their existence at different oxidation states with specific conducting properties by electronic or protonic doping.~\cite{ray1989polyaniline} 
Different forms are obtained by varying the average oxidation state and the degree of protonation ~\cite{khalil2001shear} according to the general formula~\cite{quillard1994vibrational}
\begin{equation}
    \footnotesize{ \{[-(\ce{C6H6})-\ce{NH}-(\ce{C6H6})-\ce{NH}-]_{1-\rm{x}}[-(\ce{C6H6})-\ce{N}=(\ce{C6H4})=\ce{N}-]_{\rm{x}}} \} _{\rm n}.
\end{equation}
In the above equation, $\rm{n}$ indicates the unit length of the polymer chain ({\rm n}=1 corresponds to the tetramer, {\rm n}=2 to the octamer, etc.), and $\rm{x}$ denotes an average degree of oxidation.
The latter can be varied from one to zero to give the completely reduced or the fully oxidized forms, respectively. 
The fully reduced, unprotonated form $\rm{x}=0$ is called leucoemeraldine base (LB), the half-oxidized form $\rm{x}=0.5$ emeraldine base (EB), and the fully oxidized form $\rm{x}=1$ pernigraniline base (PNB).
Their molecular structures are depicted in Figure~\ref{fig:PANI-lewis-structures}.
We should stress that the conductivity of the bare EB is not large but can be increased from about $10^{-10}$ to over 1 S/cm through, for example,  protonation in aqueous acid solutions.~\cite{macdiarmid1987polyaniline}
In such conditions, the electronic structure of PANIs is significantly altered without changing the total number of electrons in the polymer chain. 
Such features make PANIs ideal candidates for theoretical investigations.~\cite{pani-dft-lim-jcp-2000}

Experimental studies related to PANIs and their derivatives are the primary source of information on their structural, physical, and chemical characteristics.~\cite{genies1990polyaniline, quillard1994vibrational, PANI-crystal-structure}
This includes, among other things, chemical, electrochemical and gas-phase preparations, redox and polymerization mechanisms, and examinations of chemical, physical, electrochemical properties, and molecular structures.~\cite{pani-structures-x-ray, pani-moon-structural-1998, genies1990polyaniline}
Further modifications and tuning of PANI-based materials with desired properties could greatly benefit from reliable quantum chemical predictions.   
Unfortunately, such studies are limited due to computational difficulties.
It is well-known that such systems bear a non-negligible amount of multi-reference character, but their molecular size prohibits standard multi-configurational methods.
Despite that, several attempts have been made to model the electronic structures of PANIs using quantum chemistry.
One of the earliest applications is a quantum-chemical prediction of optical absorption spectra of some model PANI compounds using the intermediate neglect of differential overlap (INDO) model combined with the configuration interaction (CI) approach.~\cite{sjogren1988electronic} 
The authors were among the first to notice the importance of the torsion angle between the quinoid rings and the \ce{C-N-C} backbone. 
Semi-empirical methods were also used to study the hydration, stacking, and solvent effects of PANIs.~\cite{pani-chem-phys-2000, pani-zhekova-clusters-ijqc-2007} 
Moreover, simplified model systems of PANI were studied using Density Functional Approximations (DFAs).~\cite{pani-synth-mat-1999, pani-forman-jcpa-2003}
The Hartree--Fock (HF) and DFA optimized structures of PANIs at different oxidation states and unit lengths were investigated by Lim \textit{et al.},~\cite{pani-dft-lim-jcp-2000} Mishra \textit{et al.},~\cite{pani-dft-jpcb-2009} and Romanova~\textit{et al.}~\cite{romanova2010theoretical} 
The aforementioned studies point to an HF failure, incorrectly distributing conjugation along the polymer chain and contradicting the X-ray experimental findings.~\cite{pani-structures-x-ray} 
However, the resulting properties strongly depend on the choice of the exchange--correlation (xc) functional.
Mishra and Tandom~\cite{pani-dft-jpcb-2009} used DFAs to investigate the infrared (IR) and Raman spectra of LB and its oligomers.
Zhang \textit{et al.}.~\cite{pani-tddft} studied electronically excited states of model PANI complexes with water using time-dependent DF Theory (TD-DFT).

In this work, we reexamine the electronic structures and properties of PANIs using various approximations to the xc functional and unconventional electron correlation methods based on the pair Coupled Cluster Doubles (pCCD) model,~\cite{limacher_2013, tamar-pcc, tecmer2022geminal, geminal-review-pccp-2022} initially introduced as the Antisymmetric Product of 1-reference orbital Geminal (AP1roG) ansatz.~\cite{limacher_2013} 
An additional advantage of pCCD-based methods is the possibility to optimize all orbitals at the correlated level and a quantitative description of orbital-based correlations using concepts from quantum information theory.~\cite{ijqc-erratum, boguslawski2015orbital}
The pCCD model combined with an orbital optimization protocol~\cite{oo-ap1rog, ps2-g, pccd-orbital-rotations-jctc-2014} proved to be a reliable tool for modeling complex electronic structures and potential energy surfaces featuring strong correlation.~\cite{pawel_jpca_2014, ps2-g, pawel_pccp2015, pawel-yb2} 
Extensions to excited states within the Equation of Motion (EOM) formalism~\cite{rowe-eom, bartlett-eom} allow us to model double electron excitations,~\cite{eom-pccd, eom-pccd-erratum, eom-pccd-lccsd} a known struggle for standard EOM-CCSD-based approaches.~\cite{failure-eom-ccsd}
Furthermore, pCCD-based models allow us to gain qualitative insights into electronic structures and scrutinize them using localized orbitals, reflecting the intuitive picture of Lewis structures as pairs of electrons present from the early days of quantum chemistry.
Specifically, working in a localized basis, we will be able to unambiguously dissect electronic excitations into different types, for instance, local or charge-transfer ones.
All these features are desired in quantum chemical descriptions of electronic structures and properties of conducting polymers.  
Thus, pCCD-based quantum chemistry methods are promising alternatives to DFAs which might significantly speed up the structure-to-properties search in organic electronics and guide the experimental synthesis of new conductive polymers.

This work is organized as follows.
Section~\ref{sec:comput-det} summarizes the computational methodology.
Section~\ref{sec:results} scrutinizes the ground- and electronic excited-states properties of selected PANIs combined with a quantum information analysis of orbital correlations.
We conclude in Section~\ref{sec:conclusions}. 

\section{Computational details}\label{sec:comput-det}
\subsection{DFT calculations}
All structure optimizations and vibrational frequency calculations were performed with the Turbomole6.6~\cite{turbomole_original, turbomole} software package using the BP86~\cite{perdew86, becke88} xc functional and the def2-TZVP basis set.~\cite{turbomole-def2-tzvp, weigend1998m} The optimized xyz structures are provided in Tables S1--S7 of the ESI$\dag$.
These structures were later used for the calculation of electronic excitation energies within the TD-DFT~\cite{runge_gross, td-dft-implementation-adf} framework using the Amsterdam Density Functional (v.2018) program package,~\cite{adf1, adf2018} the BP86,~\cite{perdew86, becke88} PBE,~\cite{pbex} PBE0,~\cite{PBE0} and CAM-B3LYP~\cite{cam-b3lyp} xc functionals, and the Triple-$\zeta$ Polarization (TZ2P) basis set.~\cite{adf_b}
\subsection{pCCD-based methods}
All pCCD~\cite{geminal-review-pccp-2022, limacher_2013, oo-ap1rog, tamar-pcc} calculations were carried out in a developer version of the \textsc{PyBEST} software package~\cite{pybest-paper, pybest_zenodo} using the cc-pVDZ basis set~\cite{basis_dunning} and the DFT optimized structures.  
For the ground-state pCCD calculations, we employed the variational orbital optimization protocol.~\cite{oo-ap1rog, ps2-g, pccd-orbital-rotations-jctc-2014} 
The Pipek--Mezey localized orbitals~\cite{pipek1989fast} were used as a starting point for orbital optimization. 
Our numerical experience showed that using localized orbitals accelerates the orbital optimization process as the final pCCD natural orbitals are typically localized and bear some resemblance with split-localized orbitals.~\cite{split-localized-2003}
\subsubsection{Entanglement and correlation measures}
The 1- and 2-reduced density matrices~\cite{ijqc-2015, ijqc-erratum, entanglement_bonding_2013, post-pccd-entanglement} from variationally optimized pCCD wavefunctions were used to calculate the single orbital entropy and orbital-pair mutual information.~\cite{rissler2006, ors_ijqc, barcza_11, Ding2020, qit-concepts-schilling-jctc-2021}
The single-orbital entropies $s_i$ are calculated as~\cite{rissler2006}
\begin{equation}\label{eq:s1}
s_i=-\sum_{\alpha=1}^4 \omega_{\alpha,i} \ln \omega_{\alpha,i},
\end{equation}
where $\omega_{\alpha,i}$ are the eigenvalues of the one-orbital reduced density matrix, $\rho_{i,i'}^{(1)}$, of orbital $i$.~\cite{rissler2006, entanglement_bonding_2013, ijqc-2015, ijqc-erratum}
In the case of pCCD, such a one-orbital reduced density matrix (RDM) is 
determined from 1- and 2-particle RDMs.\cite{pccd-prb-2016, post-pccd-entanglement}
The (orbital-pair) mutual information $I_{i|j}$ is expressed as 
the difference between the amount of quantum information encoded in 
the two one-orbital reduced density matrices $i$ and $j$ and the 
two-orbital reduced density matrix associated with those two orbitals 
(the orbital pair $i,j$)~\cite{rissler2006}
\begin{equation}\label{eq:I_ij}
I_{i|j} = s_i + s_j + \sum_{\alpha=1}^{16} \omega_{\alpha,i,j} \mathrm{ln} \omega_{\alpha,i,j}.
\end{equation}
where $\omega_{\alpha,i,j}$ stands for the eigenvalues of the two-orbital 
RDM. 
Its matrix elements can be determined by generalizing the two-orbital analog of $\rho_{i,i'}^{(1)}$.~\cite{entanglement_bonding_2013, ijqc-2015, ijqc-erratum, ors_ijqc,pccd-prb-2016, post-pccd-entanglement}
The mutual information $I_{i|j}$ includes classical and quantum effects.
The classical effects usually dominate.~\cite{qit-concepts-schilling-jctc-2021}
\subsubsection{Electronic excitation energies}
The vertical electronic excitation energies were calculated using the CIS, EOM-pCCD, EOM-pCCD+S, and EOM-pCCD-CCS methods~\cite{eom-pccd, eom-pccd-erratum, pccd-ci} available in \textsc{PyBEST}.~\cite{pybest-paper, pybest_zenodo}
While in the EOM-pCCD approach, only electron-pair excitations are present in the linear excitation operator, EOM-pCCD+S and EOM-pCCD-CCS also include single excitations (see refs.~\citenum{eom-pccd, eom-pccd-erratum} for more details).
Thus, with the EOM-pCCD model, only electron-pair excitations are computed, while the EOM-pCCD+S and EOM-pCCD-CCS models allow us to determine single and double electron excitations. 
All EOM-pCCD+S calculations used the ground-state orbital-optimized pCCD reference, and all others the canonical HF orbitals. 
\section{Results and discussion}\label{sec:results}
In the following, we discuss the structural, vibrational, and electronically excited-state parameters of the aniline binary compound and selected PANIs in their tetramer and octamer structural arrangements.
Since we aim at elucidating the structure-to-property relationship in polyanilines, we require chemically reasonable structures.
In the following, we show that DFT indeed allows us to obtain reliable molecular structures, which are then used to model electronically excited states and selected properties.
The results are compared to experiments and other theoretical predictions.  
Furthermore, the TD-DFT excitation energies obtained from different xc functionals are compared to wave-function calculations. 
Finally, we use an orbital entanglement and correlation analysis of orbital interactions for assessing the electronic structures and changes in electron correlation effects in PANIs of various oxidation states and lengths. 
\begin{table*}
\small
    \caption{BP86 optimized structural parameters of PANI. Lewis structures are depicted in Figure~\ref{fig:lewis-structures-tetramer-and-octamer}, atomic labels correspond to those in Figures S1 and S2 of the ESI\dag.}
    \label{tbl:dft-structural-parameters}
    \begin{tabular*}{\textwidth}{@{\extracolsep{\fill}}llll}
    \hline
    %
    %
    \multicolumn{2}{c}{\onet} &  \multicolumn{2}{c}{\oneo} \\
    Geometrical parameters & Bond length [\AA{}] & Geometrical parameters & Bond length [\AA{}] \\ 
    \ce{N}7\ce{-C}4, \ce{N}7\ce{-C}8    & 1.393, 1.406  & \ce{N}58\ce{-C}55, \ce{N}58\ce{-C}59& 1.393, 1.404 \\
    \ce{C}4\ce{-C}3, \ce{C}8\ce{-C}9    & 1.409, 1.406  & \ce{C}55\ce{-C}54, \ce{C}59\ce{-C}60& 1.409, 1.406 \\
    \ce{C}3\ce{-C}2, \ce{C}9\ce{-C}10   & 1.395, 1.390  & \ce{C}54\ce{-C}53, \ce{C}60\ce{-C}61& 1.395, 1.391 \\
    \ce{C}2\ce{-C}1, \ce{C}10\ce{-C}11  & 1.397, 1.409  & \ce{C}53\ce{-C}52, \ce{C}61\ce{-C}62& 1.397, 1.408 \\
    \ce{C}1\ce{-C}6, \ce{C}11\ce{-C}12  & 1.399, 1.407  & \ce{C}52\ce{-C}57, \ce{C}62\ce{-C}63& 1.399, 1.408 \\
    \ce{C}6\ce{-C}5, \ce{C}12\ce{-C}13  & 1.392, 1.393  & \ce{C}57\ce{-C}56, \ce{C}63\ce{-C}64& 1.392, 1.392 \\
    \ce{C}5\ce{-C}4, \ce{C}13\ce{-C}8   & 1.411, 1.405  & \ce{C}56\ce{-C}55, \ce{C}64\ce{-C}59& 1.411, 1.406 \\
    \\
    Geometrical parameters & Bond angle [$^\circ$]  & Geometrical parameters & Bond angle [$^\circ$]  \\ 
    \ce{C}4\ce{-N}7\ce{-C}8 & 129.1 &\ce{C}55\ce{-N}58\ce{-C}59& 129.4\\
    \ce{N}7\ce{-C}4\ce{-C}3 & 123.1 &\ce{N}58\ce{-C}55\ce{-C}54& 123.0 \\
    \ce{N}7\ce{-C}8\ce{-C}9 & 122.8 &\ce{N}58\ce{-C}59\ce{-C}60& 123.1 \\
    \\
    Geometrical parameters & Dihedral angle [$^\circ$]  & Geometrical parameters & Dihedral angle [$^\circ$]  \\
    \ce{C}8\ce{-N}7\ce{-C}4\ce{-C}3 &15.1 &\ce{C}59\ce{-N}58\ce{-C}55\ce{-C}54& 22.6\\
    \ce{C}4\ce{-N}7\ce{-C}8\ce{-C}9 &36.3 &\ce{C}55\ce{-N}58\ce{-C}59\ce{-C}60& 28.1\\
    \hline
    %
    %
    \multicolumn{2}{c}{\twot} &  \multicolumn{2}{c}{\twoo} \\
    Geometrical parameters & Bond length [\AA{}] & Geometrical parameters & Bond length [\AA{}] \\ 
    \ce{N}7\ce{-C}4, \ce{N}7\ce{-C}8    &1.398, 1.394  & \ce{N}58\ce{-C}55, \ce{N}58\ce{-C}59& 1.397, 1.399 \\
    \ce{C}4\ce{-C}3, \ce{C}8\ce{-C}9    &1.408, 1.411  & \ce{C}55\ce{-C}54, \ce{C}59\ce{-C}60& 1.408, 1.407 \\
    \ce{C}3\ce{-C}2, \ce{C}9\ce{-C}10   &1.395, 1.386  & \ce{C}54\ce{-C}53, \ce{C}60\ce{-C}61& 1.395, 1.391 \\
    \ce{C}2\ce{-C}1, \ce{C}10\ce{-C}11  &1.397, 1.416  & \ce{C}53\ce{-C}52, \ce{C}61\ce{-C}62& 1.397, 1.407 \\
    \ce{C}1\ce{-C}6, \ce{C}11\ce{-C}12  &1.398, 1.419  & \ce{C}52\ce{-C}57, \ce{C}62\ce{-C}63& 1.398, 1.406 \\
    \ce{C}6\ce{-C}5, \ce{C}12\ce{-C}13  &1.392, 1.387  & \ce{C}57\ce{-C}56, \ce{C}63\ce{-C}64& 1.392, 1.391 \\
    \ce{C}5\ce{-C}4, \ce{C}13\ce{-C}8   &1.409, 1.411  & \ce{C}56\ce{-C}55, \ce{C}64\ce{-C}59& 1.410, 1.408 \\
    \\
    Geometrical parameters & Bond angle [$^\circ$]  & Geometrical parameters & Bond angle [$^\circ$]  \\ 
    \ce{C}4\ce{-N}7\ce{-C}8 & 129.9 &\ce{C}55\ce{-N}58\ce{-C}59& 129.6\\
    \ce{N}7\ce{-C}4\ce{-C}3 & 122.9 &\ce{N}58\ce{-C}55\ce{-C}54& 123.1 \\
    \ce{N}7\ce{-C}8\ce{-C}9 & 123.3 &\ce{N}58\ce{-C}59\ce{-C}60& 123.1 \\
    \\
    Geometrical parameters & Dihedral angle [$^\circ$]  & Geometrical parameters & Dihedral angle [$^\circ$]  \\
    \ce{C}8\ce{-N}7\ce{-C}4\ce{-C}3 &25.5 &\ce{C}59\ce{-N}58\ce{-C}55\ce{-C}54& 19.5\\
    \ce{C}4\ce{-N}7\ce{-C}8\ce{-C}9 &22.2&\ce{C}55\ce{-N}58\ce{-C}59\ce{-C}60& 29.7\\
    \hline
    %
    %
    \multicolumn{2}{c}{\threet} &  \multicolumn{2}{c}{\threeo} \\
    Geometrical parameters & Bond length [\AA{}] & Geometrical parameters & Bond length [\AA{}] \\ 
    \ce{N}7\ce{-C}4, \ce{N}7\ce{-C}8    &1.389, 1.313  & \ce{N}58\ce{-C}55, \ce{N}58\ce{-C}59& 1.388, 1.314 \\
    \ce{C}4\ce{-C}3, \ce{C}8\ce{-C}9    &1.414, 1.457  & \ce{C}55\ce{-C}54, \ce{C}59\ce{-C}60& 1.415, 1.457 \\
    \ce{C}3\ce{-C}2, \ce{C}9\ce{-C}10   &1.394, 1.357  & \ce{C}54\ce{-C}53, \ce{C}60\ce{-C}61& 1.394, 1.358 \\
    \ce{C}2\ce{-C}1, \ce{C}10\ce{-C}11  &1.398, 1.454  & \ce{C}53\ce{-C}52, \ce{C}61\ce{-C}62& 1.398, 1.453 \\
    \ce{C}1\ce{-C}6, \ce{C}11\ce{-C}12  &1.400, 1.456  & \ce{C}52\ce{-C}57, \ce{C}62\ce{-C}63& 1.400, 1.455 \\
    \ce{C}6\ce{-C}5, \ce{C}12\ce{-C}13  &1.391, 1.357  & \ce{C}57\ce{-C}56, \ce{C}63\ce{-C}64& 1.391, 1.358 \\
    \ce{C}5\ce{-C}4, \ce{C}13\ce{-C}8   &1.413, 1.455  & \ce{C}56\ce{-C}55, \ce{C}64\ce{-C}59& 1.413, 1.455 \\
    \\
    Geometrical parameters & Bond angle [$^\circ$]  & Geometrical parameters & Bond angle [$^\circ$]  \\ 
    \ce{C}4\ce{-N}7\ce{-C}8 &123.4 &\ce{C}55\ce{-N}58\ce{-C}59& 123.4\\
    \ce{N}7\ce{-C}4\ce{-C}3 & 123.4 &\ce{N}58\ce{-C}55\ce{-C}54& 123.4 \\
    \ce{N}7\ce{-C}8\ce{-C}9 & 123.4&\ce{N}58\ce{-C}59\ce{-C}60& 126.4 \\
    \\
    Geometrical parameters & Dihedral angle [$^\circ$]  & Geometrical parameters & Dihedral angle [$^\circ$]  \\
    \ce{C}8\ce{-N}7\ce{-C}4\ce{-C}3 &48.1 &\ce{C}59\ce{-N}58\ce{-C}55\ce{-C}54& 47.4\\
    \ce{C}4\ce{-N}7\ce{-C}8\ce{-C}9 &10.7 &\ce{C}55\ce{-N}58\ce{-C}59\ce{-C}60& 11.1\\
 \hline
 \end{tabular*}
\end{table*}

     \begin{figure}[tb]
       \centering
        \includegraphics[height=6cm]{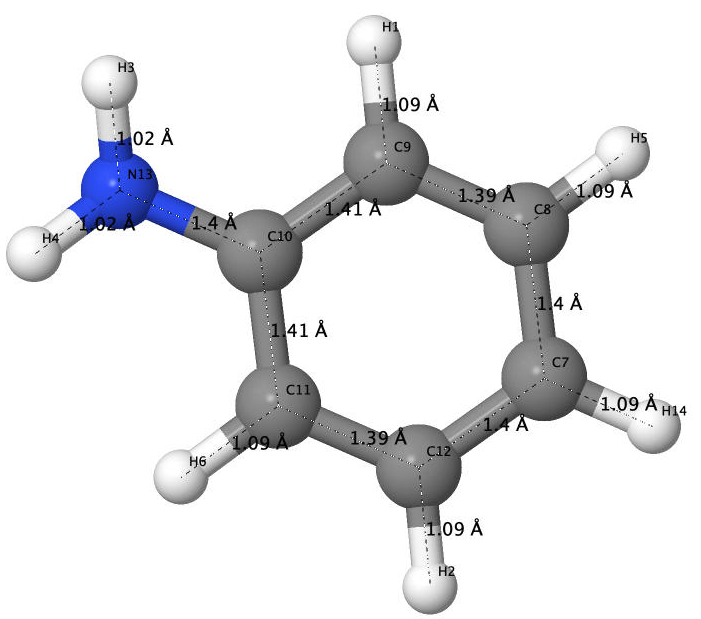}
        \caption{DFT-optimized structure of aniline including bond lengths in \AA{}.}
        \label{fig:dft-structure-aniline}
   \end{figure}

    \begin{figure*}
      \centering
       \includegraphics[height=6cm]{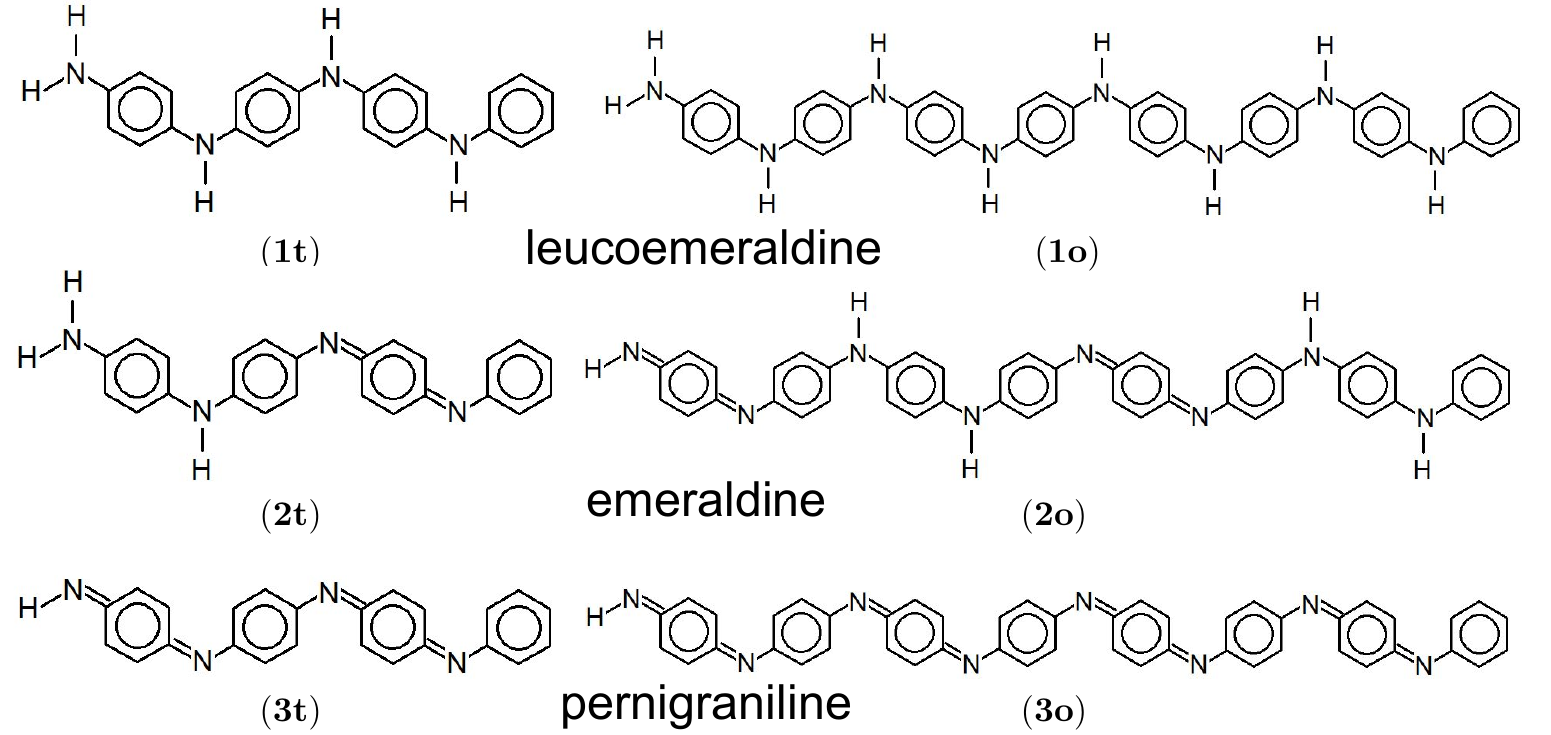}
       {\tiny}
       \caption{Lewis structures of polyanilies drawn with ChemPlot~\cite{chemplot}. Subfigures display leucoemeraldine ({\bf{1}}), emeraldine ({\bf{2}}), and pernigraniline ({\bf{3}}) in the tetramer ({\bf{t}}) and octamer ({\bf{o}}) forms.}
       \label{fig:lewis-structures-tetramer-and-octamer}
   \end{figure*}

    \begin{figure}[tb]
       \centering
       \includegraphics[height=1.8cm]{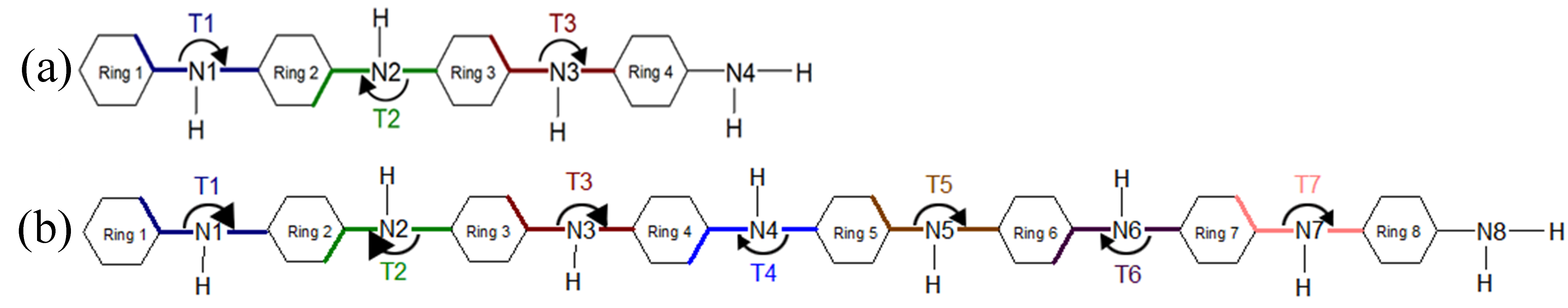}
        {\tiny}
        \caption{Schematic representation of the T1--T7 ring tilt angles in the tetramer (a) and octamer (b) models of PANIs.}
        \label{fig:tilt-angles}
    \end{figure}

\begin{table}[tb]
\small
    \caption{The ring tilt angles T1--T7 [$^{\circ}$] as labeled in Figure \ref{fig:tilt-angles} of the tetramer ({\bf{t}}) and octamer ({\bf{o}}) forms of PANIs optimized with BP86/def2-TZVP.}
    \label{tbl:tilt-angles}
    \begin{tabular*}{0.48\textwidth}{@{\extracolsep{\fill}}lllllll}
    \hline
     & \onet &  \oneo  
     & \twot &  \twoo  
     & \threet &  \threeo  \\
     T1     & 166.6 & 159.7    & 157.2 & 162.8    & 137.6 & 138.4 \\
     T2     & 194.9 & 206.3    & 212.9 & 204.7    & 191.7 & 192.7 \\
     T3     & 165.0 & 159.0    & 168.3 & 148.7    & 141.9 & 146.1 \\
     T4     &     & 201.9    &     & 192.4    &     & 194.3 \\
     T5     &     & 156.3    &     & 158.0    &     & 146.9 \\
     T6     &     & 205.2    &     & 206.4    &     & 193.5 \\
     T7     &     & 158.8    &     & 149.2    &     & 142.6 \\
    \end{tabular*}
\end{table}
\subsection{Ground-state optimized electronic structures}\label{sec:structures}
The optimized structure of aniline, a small building block of PANIs,  is shown in Figure \ref{fig:dft-structure-aniline}.
All optimized bond lengths between \ce{N-H} are roughly equal to 1.016 \AA{}, while the \ce{C-H} bond length equals 1.09 \AA{}.
The optimized structures of leucoemeraldine (\onet), emeraldine (\twot), and pernigraniline (\threet) in the tetramer form and in their corresponding octamer forms (\oneo, \twoo, and \threeo) are visualized in Figures S1 and S2 of the ESI\dag, respectively. 
Figure \ref{fig:lewis-structures-tetramer-and-octamer} shows the corresponding Lewis structures, highlighting that aniline is a building block of PANIs.
Our DFT calculations predict \ce{C-C} and \ce{C-N} bond distances between 1.3 and 1.4 \AA{} (see Table \ref{tbl:dft-structural-parameters}). 
The bond angle between two phenyl rings in \onet{} and \oneo{} and \twot{} and \twoo{} is almost the same and averages to 125$^\circ$. 
In contrast, the dihedral angles between the rings feature an average value of around 26.33$^\circ$.
For \threet{} and \threeo, the bond angle between the two phenyl rings is pretty similar except for the angle $\ce{N}58\ce{-C}59\ce{-C}60$, which increases to 126.4$^\circ$. 
The dihedral angles of \threet{} and \threeo{} significantly grow to 48.1$^\circ$ and 47.4$^\circ$, respectively, compared to \onet{} and \oneo{}.
The total torsion angle between the phenyl rings is one of the main factors that govern the band gaps, conjugation length, and electrical conductivity, all of which are important factors in determining the electronic properties of PANIs. 
For this purpose, we collected the tilt angles (as indicated in Figure \ref{fig:tilt-angles} and collected in Table~\ref{tbl:tilt-angles}) for all the tetramer (\textbf{t}) and octamer (\textbf{o}) forms of the investigated PANIs.
Our data suggest that the tilt angle for \threet~and \threeo~significantly decreases compared to the remaining PANI compounds. 
That coincides with the theoretically best conductive properties of \threet~and \threeo~over the remaining PANIs at lower oxidation states. 
\begin{table*}
\small
    \caption{Experimental and BP86 vibrational frequencies $[$cm$^{-1}]$ for aniline and all investigated PANIs.}
    \label{tbl:vib-spectra}
    \begin{tabular*}{\textwidth}{@{\extracolsep{\fill}}llllll}
\cline{6-6}
\Lower{Molecule}  & \Lower{Exp.\cite{quillard1994vibrational}}&\multicolumn{3}{c}{This work} \\ \cline{3-6}
&   & Freq. $[$cm$^{-1}]$ & Intensity $[$km/mol$]$ & Assignment \\ \hline
\\

Aniline 
 &1620& 1612 & 153.261 & \ce{N}\ce{-H}2 bending\\
 &1603 & 1580 &4.343 & \ce{C}\ce{-C} ring stretching \\
 &1276 & 1276& 53.756&  \ce{C}\ce{-N} stretching\\
 &1176, 1155 & 1147&1.333 &  \ce{C}\ce{-H} bending\\
%
%

& & \multicolumn{3}{c}{Tetramer (\bf{t})} \\ \cline{3-6}
\\
\onet
 &1618 & 1621 & 13.936& \ce{C}\ce{-C} ring stretching\\
 &     & 1616  &  18.881  & \ce{C}\ce{-C} ring stretching   \\
 &     & 1615  &  26.274  &  \ce{C}\ce{-C} ring stretching  \\
 &     & 1599  &  206.853  &  \ce{C}\ce{-C} ring stretching  \\
 &     & 1602 & 62.197 & \ce{N}\ce{-H}2 bending\\
 &1219 & 1221&15.378 &  \ce{C}\ce{-N} stretching \\
 &     & 1219  & 6.463   &  \ce{C}\ce{-N} stretching   \\
 &1181 & 1165& 4.293 &  \ce{C}\ce{-H} bending\\
 &     & 1163  &  0.161  &  \ce{C}\ce{-H} bending   \\
\\
\twot
 &1617 & 1619 & 91.950 & \ce{C}\ce{-C} ring stretching\\
 &     & 1606 &506.207  & \ce{N}\ce{-H}2 bending\\
 &1519 & 1519&254.253 &  \ce{C}\ce{=N} stretching\\
 &1220, 1219 & 1227&6.510 &  \ce{C}\ce{-N} stretching\\
 &    & 1222 & 3.694&  \ce{C}\ce{-N} stretching\\
 &1182 & 1168&  2.785   &  \ce{C}\ce{-H} bending\\
 &     & 1155  & 132.108   &  \ce{C}\ce{-H} bending    \\
 &     & 1153  &  150.552  &  \ce{C}\ce{-H} bending   \\
 &     & 1144  & 1.584   &  \ce{C}\ce{-H} bending   \\
\\
\threet
 &1612, 1553 & 1556 & 49.940   & \ce{C}\ce{-C} ring stretching \\
 &1582, 1579 & 1588 &  2.021  & \ce{C}\ce{=C} stretching \\ 
 &    & 1581  &  5.799  &\ce{C}\ce{=C} stretching\\
 &1480 & 1496& 77.580  &  \ce{C}\ce{=N} stretching\\
 &1219 & 1235& 0.642&  \ce{C}\ce{-N} stretching\\
 &    & 1229 &7.934 &  \ce{C}\ce{-N} stretching\\
 &    & 1218 & 22.962 &  \ce{C}\ce{-N} stretching\\
 &1157 & 1157&  2.481   &  \ce{C}\ce{-H} bending\\
%
%
& & \multicolumn{3}{c}{Octamer (\bf{o})} \\ \cline{3-6}
\\
\oneo
 &1618 & 1622 & 5.474 & \ce{C}\ce{-C} ring stretching \\
 &     & 1617  & 0.271   & \ce{C}\ce{-C} ring stretching   \\
 &     & 1616  &   5.865 &  \ce{C}\ce{-C} ring stretching  \\
 &     & 1615  &  16.793  &  \ce{C}\ce{-C} ring stretching   \\
 &     & 1599  &  206.640  &  \ce{C}\ce{-C} ring stretching   \\
 &     & 1602 & 70.742 & \ce{N}\ce{-H}2 bending\\
 &1219 & 1221& 1.003& \ce{C}\ce{-N} stretching\\
 &     & 1220  & 7.350   &  \ce{C}\ce{-N} stretching  \\
 &     & 1219  &  41.414  &  \ce{C}\ce{-N} stretching   \\
 &1181 & 1165& 6.165&  \ce{C}\ce{-H} bending\\
 &     & 1164  &  0.405  &  \ce{C}\ce{-H} bending   \\
 &     & 1163  &   1.103 &  \ce{C}\ce{-H} bending   \\
\\
\twoo
 &1617 & 1616 &23.481 & \ce{C}\ce{-C} ring stretching \\
 &1519 & 1515&169.019 &  \ce{C}\ce{=N} stretching\\
 &1220, 1219 & 1224&45.260 &  \ce{C}\ce{-N} stretching\\
 &    & 1223 &21.564 &  \ce{C}\ce{-N} stretching\\
 &    & 1221 & 4.281&  \ce{C}\ce{-N} stretching\\
 &1182 & 1169&   1.030  &  \ce{C}\ce{-H} bending\\
 &     & 1166  & 3.486   &  \ce{C}\ce{-H} bending   \\
 &     & 1165  &  76.322  &  \ce{C}\ce{-H} bending    \\
 &     & 1157  &  113.823  &  \ce{C}\ce{-H} bending   \\
 &     & 1155  &  455.013  & \ce{C}\ce{-H} bending   \\
 &     & 1149  & 0.902   &  \ce{C}\ce{-H} bending   \\
 &     & 1143  &  21.476  &  \ce{C}\ce{-H} bending   \\
\\
\threeo
 &1612, 1553 & 1586 &  8.260  & \ce{C}\ce{-C} ring stretching  \\
 &    & 1581 &  5.022  & \ce{C}\ce{-C} ring stretching  \\
 &1582, 1579 & 1589 &  5.578  & \ce{C}\ce{=C} stretching\\ 
 &    & 1574  &  7.336  & \ce{C}\ce{=C} stretching \\
 &1480 & 1490&  158.490 &  \ce{C}\ce{=N} stretching\\
 & & 1472& 0.771  &  \ce{C}\ce{=N} stretching\\
 &1219 & 1218  &  53.816 &  \ce{C}\ce{-N} stretching\\
 &1157 & 1157& 8.286    &  \ce{C}\ce{-H} bending\\
 &   & 1148&  269.192   &  \ce{C}\ce{-H} bending\\
 \end{tabular*}
\end{table*}
We should also stress that the effect of various approximate xc functionals on the torsional angle of PANI is discussed in Ref.~\citenum{romanova2010theoretical}
The studies conclude that different xc functionals provide qualitatively the same torsional angles.

\subsection{Vibrational spectra}
Aniline and PANIs have been a significant target of structural and electronic studies, experimentally and theoretically, for many years~\cite{aniline-experimental-ir, aniline-hobza, quillard1994vibrational,pani-dft-lim-jcp-2000, mishra2015dft, pani-dft-jpcb-2009}.
Table \ref{tbl:vib-spectra} presents a complete vibrational assignment of all fundamental vibrations and a comparison to experimental data.~\cite{quillard1994vibrational}
Most importantly, all theoretical data agrees with experimental results for aniline and PANIs.
The vibrational spectra of all investigated PANIs are reconstructed in Figure S3 of the ESI\dag~using the \textsc{Gabedit} software package. 
In the spectrum of aniline, two peaks appear at 1612 and 1580 cm$^{-1}$.
The former is assigned to the $\ce{-N}\ce{H}2$ bending and the latter to the $\ce{C}\ce{-C}$ ring-stretching vibration of the phenyl group.
The remaining leading vibrations of the Raman and IR spectra are located at 1276 cm$^{-1}$ and correspond to the ring-stretching mode mainly attributed to the $\ce{C}\ce{-N}$ stretching.
The band at 1147 cm$^{-1}$ results from the $\ce{C}\ce{-H}$ bending mode. 
All the characteristic features of the aniline vibrational spectrum are present in all investigated PANIs, except for the \ce{-NH2} peak that is absent in \threet~and \threeo. 

For \onet{} we observe several characteristic vibrations of the benzene ring, such as those peaked at 1599, 1615, 1616, and 1621 cm$^{-1}$, which correspond to a $\ce{C}\ce{-C}$ stretching vibrational mode for ring 1, 2, 3, and 4, respectively, (cf.~Figure \ref{fig:tilt-angles} for ring labels) and two $\ce{C}\ce{-H}$ bending vibrational modes at 1165 and 1163 cm$^{-1}$. 
The bands at 1221, and 1219 cm$^{-1}$ correspond to the $\ce{C}\ce{-N}$ stretching vibrational mode for $\ce{N}1$, $\ce{N}2$, and $\ce{N}3$ respectively, while the $\ce{-N}\ce{H2}$ bending mode is positioned at 1602 cm$^{-1}$ (the atomic labels are indicated in Figure S1 of the ESI\dag). 

For \twot, the $\ce{C}\ce{-C}$ ring-stretching is located at 1619 cm$^{-1}$, and the $\ce{C}\ce{=N}$ stretching mode at 1519 cm$^{-1}$.
The two peaks at 1222 ($\ce{N}1$) and 1227 ($\ce{N}2$ and $\ce{N}3$) cm$^{-1}$ are due to a $\ce{C}\ce{-N}$ stretching mode
(see also Figure S1 of the ESI\dag~for atomic labels).
The $\ce{C}\ce{-H}$ bending vibrational mode of the benzene ring can be characterized by a Raman band at 1168, 1155, 1153, and 1144 cm$^{-1}$, respectively.
The $\ce{-N}\ce{H2}$ bending mode is positioned at 1606 cm$^{-1}$.

\threet{} features the fundamental bands of $\ce{C}\ce{=C}$ stretching modes at 1581 and 1588 cm$^{-1}$ and a $\ce{C}\ce{-C}$ ring-stretching mode at 1556 cm$^{-1}$.
The Raman band at 1496 cm$^{-1}$ corresponds to a $\ce{C}\ce{=N}$ stretching vibrational mode, while the $\ce{C}\ce{-N}$ stretching mode is positioned at 1217, 1228, and 1234 cm$^{-1}$. The $\ce{C}\ce{-H}$ bending mode is predicted at 1157 cm$^{-1}$.

Comparing the characteristic vibrational features of \onet, \twot, and \threet{}, we note a redshift of the \ce{C-C} ring stretching and \ce{C-H} bending frequencies. 
Moreover, we observe a blueshift of the \ce{N-H2} bending vibrations from \onet{} to \twot{}.
Essentially the same vibrational features as for \onet, \twot, and \threet{} are observed for \oneo, \twoo, and \threeo, respectively. 
The only difference is the larger number of peaks and a negligible increase in characteristic vibrational frequencies by about 1-2 cm$^{-1}$ for longer polymer chains (cf.~Table~\ref{tbl:vib-spectra}). 
 
\subsection{HOMO--LUMO gaps from DFAs}
The HOMO and LUMO molecular orbitals of \onet, \twot, \threet, \oneo, \twoo, and \threeo~ obtained from different xc functionals (BP86, PBE, PBE0, and CAM-B3LYP) are depicted in Figures S5-S8 of the ESI\dag. 
All xc functionals predict similar HOMO and LUMO $\pi$- and $\pi^*$-type molecular orbitals delocalized over the whole molecular structures. 
The HOMO and LUMO energies and the HOMO--LUMO gaps are summarized in Table S8 and visualized in Figure S4 of the ESI\dag.
Both generalized gradient approximations to the xc functional (BP86 and PBE) predict identical HOMO--LUMO gaps for aniline and almost identical for all PANIs. 
The PBE0 xc functional with an admixture of 25${\%}$ of HF exchange roughly doubles the HOMO--LUMO gaps. 
The range-separated CAM-B3LYP xc functional further widens the HOMO--LUMO gaps by about 20-25${\%}$. 
Specifically, CAM-B3LYP predicts the HOMO--LUMO gap of 0.29 eV for aniline, and 0.196 eV for \onet, 0.155 eV for \twot, and 0.162 eV for \threet, respectively.
The HOMO--LUMO gap is only slightly affected (lowered by around 0.01 eV) in the longer PANIs (\oneo, \twoo, and \threeo). 
Finally, we should note that our DFA calculations do not show any clear trend of the HOMO--LUMO gap with respect to the formal oxidation state of PANIs. 


\begin{table*}
\small
    \caption{Lowest-lying singlet--singlet excitation energies [eV] and their characteristics calculated from TD-DFT, EOM-pCCD+S, and CIS. HOMO and LUMO are abbreviated  as H and L, respectively. Note that EOM-pCCD+S was performed using the natural pCCD orbitals, which are ordered with respect to occupation numbers, not orbital energies. For EOM-pCCD+S, only the leading contribution is shown. Table~\ref{tbl:collective-electronic-excitations} dissects each excited state with respect to selected transitions.}
    \label{tbl:electronic-excitations}
    \centering
    \resizebox{0.81\linewidth}{!}{%
    \begin{tabular*}{\textwidth}{@{\extracolsep{\fill}}lllllllll}
Molecule  & no. & character& BP86 & PBE & PBE0 & CAM-B3LYP & EOM-pCCD+S & CIS  \\ \hline

\multirow{9}{*}{\STAB{\rotatebox[origin=c]{90}{Aniline}}} 

 && energy & 4.422 & 4.410 &4.820&4.912&6.005&5.821\\
 & & weight &0.900&0.900&0.880&0.860&0.485&0.618 \\
 &1 & character&  H$\rightarrow$ L&  H$\rightarrow$ L& H$\rightarrow$ L & H$\rightarrow$ L& H-1$\rightarrow$ L& H$\rightarrow$ L\\
 & & intensity&0.029 & 0.029&0.038 &0.040&--&-- \\
 \cline{2-9}
 && energy &4.903&4.660&5.180&5.253&6.880&6.174\\
& & weight &0.850&0.810&0.550&0.950&0.445&0.551 \\
 &2 & character&  H$\rightarrow$ L+2&  H$\rightarrow$ L+2& H$\rightarrow$ L+2& H$\rightarrow$ L+1& H-1$\rightarrow$ L+1& H$\rightarrow$ L+1\\
 & & intensity&0.013 &0.008&0.013&0.012&--&--\\
 \cline{2-9}
&& energy &5.373 &5.250&5.670&5.737&8.002&7.304\\
& & weight &0.660&0.820&0.430&0.810&0.309&0.578\\
 &3 & character& H$\rightarrow$ L+1& H$\rightarrow$ L+3& H$\rightarrow$ L+1& H$\rightarrow$ L+2& H-11$\rightarrow$ L+3& H$\rightarrow$ L+2\\
 & & intensity&0.131&0.024&0.130&0.128&--&--\\
 \hline 

\multirow{9}{*}{\STAB{\rotatebox[origin=c]{90}{\onet}}} 
&& energy & 2.791 & 2.780 &3.547&3.916&5.525&4.913\\
 & & weight &0.570&0.490&0.720&0.600&0.192&0.552 \\
 &1 & character&H$\rightarrow$ L&  H$\rightarrow$ L& H$\rightarrow$ L& H$\rightarrow$ L& H-38$\rightarrow$ L+3& H$\rightarrow$ L\\
 & & intensity&0.027&0.025&0.562&0.839&--&-- \\
 \cline{2-9}
 && energy &2.834&2.823&3.632&4.000&5.645&5.116\\
& & weight &0.720&0.770&0.690&0.530&0.231&0.534 \\
 &2 & character& H$\rightarrow$ L+2& H$\rightarrow$ L+2& H$\rightarrow$ L+1& H$\rightarrow$ L+1& H-32$\rightarrow$ L+3& H$\rightarrow$ L+1\\
 & & intensity&0.036&0.035&0.724&0.558&--\\
 \cline{2-9}
&& energy &2.918&2.894&3.700&4.092&5.692&5.220\\
& & weight &0.440&0.430&0.880&0.600&0.293&0.484 \\
 &3 & character&H$\rightarrow$ L+1&  H$\rightarrow$ L+1&H$\rightarrow$ L+2& H$\rightarrow$ L+2& H-33$\rightarrow$ L+1& H$\rightarrow$ L+2\\
 & & intensity&0.647&0.671&0.033&0.032&--&--\\
 \hline 
\multirow{9}{*}{\STAB{\rotatebox[origin=c]{90}{\twot}}} 
&& energy & 1.734&1.729&2.078&2.376&4.381&3.173\\
 & & weight &0.880&0.880&0.940&0.890&0.212&0.628 \\
 &1 & character&H$\rightarrow$ L&  H$\rightarrow$ L& H$\rightarrow$ L& H$\rightarrow$ L& H-43$\rightarrow$ L+1& H$\rightarrow$ L\\
 & & intensity&0.912&0.909&1.177&1.301&--&--\\
 \cline{2-9}
 && energy &2.012&2.001&2.429&2.814&4.987&3.872\\
& & weight &0.890&0.890&0.890&0.740&0.332&0.509 \\
 &2 & character& H-1$\rightarrow$ L& H-1$\rightarrow$ L& H-1$\rightarrow$ L& H-1$\rightarrow$ L& H-42$\rightarrow$ L& H-1$\rightarrow$ L\\
 & & intensity&0.091&0.087&0.022&0.001&--&--\\
 \cline{2-9}
&& energy &2.583&2.573&3.304&3.896&5.821&4.972\\
& & weight &0.890&0.900&0.610&0.190&0.312&0.429 \\
 &3 & character& H-2$\rightarrow$ L&  H-2$\rightarrow$ L& H-2$\rightarrow$ L& H-4$\rightarrow$ L& H-28$\rightarrow$ L+6& H-9$\rightarrow$ L\\
 & & intensity&0.001&0.001&0.021&0.674&--&--\\
 \hline 
 \multirow{9}{*}{\STAB{\rotatebox[origin=c]{90}{\threet}}} 

&& energy &1.676 &1.671&2.087&2.413&4.468&3.216\\
 & & weight &0.580&0.580&0.940&0.870&0.255&0.617 \\
 &1 & character& H$\rightarrow$ L& H$\rightarrow$ L& H$\rightarrow$ L& H$\rightarrow$ L& H-45$\rightarrow$ L+2& H$\rightarrow$ L\\
 & & intensity&0.566&0.564&1.198&1.327&--&-- \\
 \cline{2-9}
 && energy &1.770&1.763&2.450&2.916&5.123&3.923\\
& & weight &0.380&0.370&0.500&0.440&0.384&0.405 \\
 &2 & character& H$\rightarrow$ L& H$\rightarrow$ L& H$\rightarrow$ L+1&H-1$\rightarrow$ L& H-46$\rightarrow$ L+1& H-1$\rightarrow$ L\\
 & & intensity&0.378&0.367&0.010&0.025&--&--\\
 \cline{2-9}
&& energy &2.012&2.002&2.539&2.942&5.284&4.049\\
& & weight &0.540&0.560&0.440&0.430&0.382&0.343 \\
 &3 & character& H$\rightarrow$ L+1& H$\rightarrow$ L+1& H$\rightarrow$ L+1& H$\rightarrow$ L+1& H-45$\rightarrow$ L+2& H$\rightarrow$ L+1\\
 & & intensity&0.005&0.005&0.011&0.021&--&--\\

 \hline 

\multirow{9}{*}{\STAB{\rotatebox[origin=c]{90}{\oneo}}} 
&& energy & 2.381&2.364&3.266&${^*}$--&5.349&4.683\\
 & & weight &0.900&0.900&0.670&${^*}$--&0.111&0.408 \\
 &1 & character& H$\rightarrow$ L&  H$\rightarrow$ L&H$\rightarrow$ L&${^*}$--&H-78$\rightarrow$ L+9& H$\rightarrow$ L\\
 & & intensity&0.375&0.397&2.618&${^*}$--&--&-- \\
 \cline{2-9}
 && energy &2.473&2.456&3.470&${^*}$--&5.516&4.899\\
& & weight &0.870&0.900&0.280&${^*}$--&0.165&0.342 \\
 &2 & character& H$\rightarrow$ L+1&H$\rightarrow$ L+1& H$\rightarrow$ L+4&${^*}$--&H-53$\rightarrow$ L+6& H-1$\rightarrow$ L\\
 & & intensity&0.085&0.061&0.111&${^*}$--&--&--\\
 \cline{2-9}
&& energy &2.535&2.523&3.486&${^*}$--&5.548&5.043\\
& & weight &0.570&0.530&0.370&${^*}$--&0.179&0.267 \\
 &3 & character&H$\rightarrow$ L+2& H$\rightarrow$ L+2& H$\rightarrow$ L+3&${^*}$--&H-52$\rightarrow$ L+3& H$\rightarrow$ L+2\\
 & & intensity&0.007&0.009&0.253&${^*}$--&--&--\\
 \hline 
\multirow{9}{*}{\STAB{\rotatebox[origin=c]{90}{\twoo}}} 
&& energy &0.970&0.970&1.783&${^*}$--&4.249&3.042\\
 & & weight &0.780&0.770&0.940&${^*}$--&0.204&0.559 \\
 &1 & character&  H$\rightarrow$ L&  H$\rightarrow$ L& H$\rightarrow$ L&${^*}$--&H-85$\rightarrow$ L+1& H$\rightarrow$ L\\
 & & intensity&0.150&0.152&2.217&${^*}$--&--&-- \\
 \cline{2-9}
 && energy &1.272&1.268&1.996&${^*}$--&4.675&3.548\\
& & weight &0.410&0.400&0.950&${^*}$-- &0.329&0.374\\
 &2 & character& H$\rightarrow$ L+1& H$\rightarrow$ L+1& H$\rightarrow$ L+1&${^*}$--&H-84$\rightarrow$ L+2& H-1$\rightarrow$ L+1\\
 & & intensity&1.379&1.386&0.002&${^*}$--&--&--\\
 \cline{2-9}
&& energy &1.347&1.345&2.088&${^*}$--&4.896&3.792\\
& & weight &0.290&0.290&0.910&${^*}$--&0.310&0.326 \\
 &3 & character&  H-1$\rightarrow$ L& H-1$\rightarrow$ L& H-1$\rightarrow$ L&${^*}$--&H-86$\rightarrow$ L+3&H-3$\rightarrow$ L\\
 & & intensity&0.013&0.010&0.004&${^*}$--&--&--\\
 \hline 
 \multirow{9}{*}{\STAB{\rotatebox[origin=c]{90}{\threeo}}} 
&& energy &1.093&1.090&1.524&${^*}$--&3.959&2.598\\
 & & weight &0.480&0.480&0.930&${^*}$--&0.159&0.558 \\
 &1 & character&  H$\rightarrow$ L+1&  H$\rightarrow$ L+1&H$\rightarrow$ L&${^*}$--&H-3$\rightarrow$ L+3& H$\rightarrow$ L\\
 & & intensity&0.090&0.069&3.561&${^*}$--&--&-- \\
 \cline{2-9}
 && energy &1.117&1.114&1.893&${^*}$--&4.370&3.135\\
& & weight &0.810&0.810&0.480&${^*}$--&0.174&0.408 \\
 &2 & character&H$\rightarrow$ L&  H$\rightarrow$ L& H-1$\rightarrow$ L&${^*}$--&H-81$\rightarrow$ L+4& H-1$\rightarrow$ L\\
 & & intensity&1.970&1.976&0.000&${^*}$--&--&--\\
 \cline{2-9}
&& energy &1.349&1.344&2.028&${^*}$--&4.679&3.524\\
& & weight &0.550&0.560&0.490&${^*}$--&0.216&0.382 \\
 &3 & character&H-2$\rightarrow$ L& H-2$\rightarrow$ L& H$\rightarrow$ L+1&${^*}$--&H-84$\rightarrow$ L+3& H-2$\rightarrow$ L\\
 & & intensity&0.003&0.009&0.000&${^*}$--&--&--\\

 \cline{1-9} 
\\
 \end{tabular*}
 }
 \begin{tablenotes}\footnotesize
    \item $^*$ The CAM-B3LYP ground-state calculations for \oneo, \twoo, and \threeo{} did not converge due to numerical difficulties.
    \end{tablenotes}

\end{table*}

\begin{table*}
\small
    \caption{Collective EOM-pCCD+S contributions to a given type of excitation in PANIs. LP$_{\rm N}$ denotes the lone pair on nitrogen, B --- benzenoid ring, Q --- quinoid ring, and $\sigma_{\rm N}$ --- sigma-type orbital in nitrogen, respectively. Note that LP$_{\rm N}$, B, Q, and $\sigma_{\rm N}$ indicate localized orbitals on each individual fragment, e.g., Q indicates orbitals centered solely on the quinoid ring. For a detailed discussion, see text.}
    \label{tbl:collective-electronic-excitations}
    \centering
    \begin{tabular*}{\textwidth}{@{\extracolsep{\fill}}cllllllll}
    
Molecule  & no. &LP$_{\rm N}\rightarrow$ B &LP$_{\rm N}\rightarrow$ Q &$\sigma_{\rm N}$ $\rightarrow$ Q&  B $\rightarrow$ B  & B $\rightarrow$ Q & Q $\rightarrow$ B & Q $\rightarrow$ Q   \\ \hline

\onet
 &1 & 12.4${\%}$ & & &38.8${\%}$ & & & \\
 &  2   & 13.7${\%}$ & & & 40.3 ${\%}$   &  & &  \\
 &  3   & 16.5 ${\%}$ & & & 51 ${\%}$  &  & & \\
\cline{2-9}
\twot
 &1 &  & 12.5${\%}$ & 1.2${\%}$& 1.5${\%}$& 25.8 ${\%}$& & 24.2${\%}$ \\
 &  2   & &29.3${\%}$ &  4.6${\%}$  &1.6${\%}$&26.5${\%}$& & 2.9${\%}$\\
 &  3   &0.64${\%}$ & 5.6${\%}$  & & 62.7${\%}$& 1.3${\%}$& 1.9${\%}$ \\
\cline{2-9}
\threet
 &  1   & &10.3${\%}$ & 0.9${\%}$ & 8.1${\%}$&19.2${\%}$& & 21.7${\%}$ \\
 &  2   & & 19.8${\%}$ &3.5${\%}$&4.3${\%}$&27.3${\%}$&& 16.4${\%}$ \\
 &  3   & & 28.3${\%}$  & 4.5${\%}$&& 21.7${\%}$&& 15.3${\%}$ \\
\hline
&&&&&&&& \\
\oneo
&1 & 9.3${\%}$ &  &  &14.1${\%}$&  &  &  \\
 &  2   & 9.3${\%}$ &  &  &32.2${\%}$&  &  &  \\
 &  3   & 11.2${\%}$ &  &  &36.8${\%}$&  &  &  \\
\cline{2-9}
\twoo
&1 &  & 10.3${\%}$ & 1.1${\%}$&1.4${\%}$&20.9${\%}$&& 25.3${\%}$\\
 &  2   &0.63${\%}$ &  16.7${\%}$  & 3.8${\%}$& 7.2${\%}$&25.9${\%}$&& 12.5${\%}$\\
 &  3   & & 25.3${\%}$  &  5.1${\%}$&1.5${\%}$&23.3${\%}$&&3.2${\%}$ \\
\cline{2-9}
\threeo
&1 &  & 7.5${\%}$& & 1.1${\%}$& 13.9${\%}$& & 19.3${\%}$\\
&2 & & 11.08${\%}$ &  & 1.7${\%}$ & 14.9${\%}$ &  & 9${\%}$\\
 &  3   & & 17.7${\%}$  & 0.6${\%}$& 1${\%}$&15${\%}$&0.5${\%}$&4.3${\%}$\\
\hline  
 \end{tabular*}
\end{table*}

\subsection{Electronic excitation energies}
A significant feature of conjugated polymers often studied theoretically and experimentally is the electronic structure of their valence band. 
The desired donor properties feature high-intensity electronic transitions with a dominant HOMO $\rightarrow$ LUMO character in the specific range of the spectrum.~\cite{cui2020recent}
Therefore, we will scrutinize the lowest-lying electronic excitation energies obtained from different quantum chemistry methods to assess the structure-to-property relationship.  
Table~\ref{tbl:electronic-excitations} summarizes low-lying electronic transition energies and associated characteristics obtained from various xc functionals (BP86, PBE, PBE0, and CAM-B3LYP), CIS, and EOM-pCCD+S. 
The EOM-pCCD and EOM-pCCD-CCS excitation energies are reported in Table S9 of the ESI\dag~for comparison. 
\subsubsection{TD-DFT and CIS excitation energies}
The HOMO $\rightarrow$ LUMO excitations dominate the first excitation energy in TD-DFT studies of all investigated molecules and have non-zero transition dipole moments (TDMs).  
The higher-lying excitations involve mainly an electron transfer from HOMO to LUMO+1 and LUMO+2 orbitals, with the latter having $\pi^*$ character.
An exception is \textbf{2}, for which the second and third excited states occur from $\pi$-type orbitals located below the HOMO.
Thus, the low-lying part of the electronic spectrum of PANIs is dominated by $\pi \rightarrow \pi^*$ transitions. 
Unfortunately, the delocalized nature of DFA orbitals prevents us from assessing the character of electronic transitions in PANIs with more details.
We should also stress that the HOMO/LUMO orbital energies and the HOMO--LUMO gaps discussed in the previous subsection do not correlate with the low-lying part of the electronic spectrum of PANIs. 

PANIs significantly lower the electronic transitions observed in the aniline model system. Specifically, they fall in an energetic descending order \onet{} < \oneo{} < \threet{} < \threeo{} < \twot{} < \twoo, indicating that emeraldine has the lowest-lying electronic transitions among them all.~\cite{genies1990polyaniline}
That contradicts the common experimental knowledge about the absorption spectra of PANI, which is expected to be in the range of 2-3.2 eV for leucoemeraldine, 1.6-3.1 eV for emeraldine, and 1.0-1.8 eV for pernigraniline, respectively.~\cite{genies1990polyaniline, sjogren1988electronic}
We should stress that the electrical conductivity of PANIs does not directly depend on the position of their excited states but on the type of dopant, the extent of doping, and the polymer length.~\cite{macdiarmid2001-noble-lecture} 
Such features combined with the given range of excitation energies and their type can, in turn, affect the conjugation properties. 

Moving from structures \textbf{t} to \textbf{o}, we observe a lowering of excitations by about 0.3-0.4 eV.
The absolute values of excitation energies and, to some extent, their characteristics strongly depend on the applied xc functional.

Based on previous TD-DFT benchmarks and analysis of excitation energies, we do not expect any outstanding performance from semi-local xc functionals like BP86 and PBE, as they tend to underestimate electron excitations.~\cite{excitations-review, td-dft-benchmark-jacquemin, dft-organic-electronics-range-separation} 
Addition of HF exchange introduces some non-local effects in the xc kernel and improves the overall performance of TD-DFT.
We expect further enhancement of the description of charge-transfer states with range-separated hybrids.~\cite{dreuw_2003, dreuw_2004, pawel1, pawel2, pawel_saldien}
Thus, we anticipate the PBE0 and CAM-B3LYP results to be more reliable, although limited to model single electronic transitions and electronic structures well-described by a single Slater determinant. 
A significant difference between the PBE0 and CAM-B3LYP excitation energies can be used to identify partial charge-transfer states.~\cite{pawel2}
Based on that, we anticipate that all investigated PANI structures have some partial charge-transfer character, with aniline being the exception.
The nature of PBE0 and CAM-B3LYP transitions is very similar, except for structure \onet{}, where the order of the 2-nd and 3-rd excited state changes.  
The PBE0 and CAM-B3LYP excitation energies are comparable in magnitude to the CIS data: electronic transitions' ordering and main character are virtually the same. 
They differ, however, in the absolute values of excitation energies (cf. Table~\ref{tbl:electronic-excitations}), where CIS predicts much higher excitation energies.
The most considerable discrepancies are observed for the aniline molecule (up to 1.5 eV) and are reduced to approximately 1 eV in PANIs. 
\\
\subsubsection{EOM-pCCD+S excitation energies}
The EOM-pCCD-based electronic spectra proved to be reliable even for complex electronic structures.~\cite{pccd-ee-f0-actinides, eom-pccd, eom-pccd-lccsd} 
Even the simplest and computationally most efficient EOM-pCCD+S model provides trustworthy results for long polymer chains.~\cite{eom-pccd, eom-pccd-lccsd}
Most importantly, for singly-excited states, EOM-pCCD+S correctly determines the main character of excitation energies, while the actual excitation energies differ within 0.1 to 0.2 eV from more elaborate EOM-CC models, even outperforming multireference methods like the density matrix renormalization group algorithm.~\cite{eom-pccd, eom-pccd-lccsd}
That motivates us to use this EOM-CC flavor for modeling the electronic spectra of PANI compounds, which are too expensive for conventional EOM-CC methods like EOM-CCSD. 
The EOM-pCCD+S excitation energies listed Table~\ref{tbl:electronic-excitations} have dominant contributions from single electronic excitations. 
Next to each EOM-pCCD+S state, the component with the largest weight is provided.
We should note that the pCCD-optimized orbitals are sorted according to their (natural) occupation numbers, whose order does not correspond to the energetic ordering of the canonical HF orbitals.
The EOM-pCCD+S results are higher than CIS by about 0.6 eV for \onet, 1.2 eV for \twot, 1.3 eV for \threet, 0.7 eV for \oneo, 1.2 eV for \twoo, and 1.4 eV for \threeo{} and overall higher by about 1-2 eV than the PBE and CAM-B3LYP results.
However, based on previous numerical experience with similar systems,~\cite{eom-pccd, eom-pccd-lccsd} we are convinced that the EOM-pCCD+S provides the correct character of excited states and can be employed for a deeper analysis.
A significant difference between the EOM-pCCD+S and CIS  methods originates from the orbital bases: EOM-pCCD+S utilizes the pCCD-optimized orbitals that are localized in nature (see Figures S10-S18 of the ESI\dag), while CIS uses the canonical HF orbitals (delocalized).
Thus, the pCCD-optimized orbitals offer a different viewpoint, in which the information is more compressed, and we have many components in electronic transitions.~\cite{local-orbitals-in-quantum-chemistry, stewart-interpretation-of-localized-orbitals-2019}
Unlike TD-DFT and CIS, where the electronic transitions are dominated by one main electronic configuration, each electronic transition in EOM-pCCD+S includes several orbital contributions of similar weights but often of various characteristics (the pCCD orbitals involved in the low-lying excitations are shown in Figures S10-S18 of the ESI\dag).
Finally, to the best of our knowledge, no reliable experimental electronic spectra exist of the PANI series studied in this work. 
Yet, the lack of environmental or crystal structure packing effects in ab initio calculations prohibits us from directly comparing theory and experiment.
\\
\textbf{Collective contributions to excitation energies. }
{Table~\ref{tbl:collective-electronic-excitations} summarizes the collective contributions to each excited state, where all the excitation contributions are grouped according to their character.
We can see the qualitative differences in the low-lying transitions between the leucoemeraldine (\onet{} and \oneo{}) and the remaining structures; 
while the lower part of the leucoemeraldine spectrum is dominated by the LP$_{\rm{N}}\rightarrow$B (N lone pair to benzenoid ring) and B$\rightarrow$B electronic excitations, in the remaining systems, the electrons are mainly transferred to the quinoid ring (Q).
Specifically, the electronic spectrum of emeraldine (\twot{} and \twoo{}) and pernigraniline (\threet{} and \threeo{}) are best described by the LP$_{\rm{N}}\rightarrow$Q, Q$\rightarrow$Q, and B$\rightarrow$Q electronic transitions. 
Their collective contributions increase for higher-lying states.
Thus, the local nature of pCCD-optimized orbitals allows us to dissect the character of each transition in PANIs and their structure-to-property relationship.
Specifically, the LP$_{\rm{N}}\rightarrow$Q transitions in leucoemeraldine can be classified as charge-transfer (CT) type and the B$\rightarrow$B as local (L) in nature. 
All B$\rightarrow$Q excitations in emeraldine and pernigraniline are of CT type, and Q$\rightarrow$Q are L type.
The LP$_{\rm{N}}\rightarrow$Q electronic transitions in emeraldine have dominantly CT character, but in pernigraniline, they have mixed CT/L nature with a diminishing CT character in the longer polymer structure (\threeo{}).
}

An additional feature of the leucoemeraldine electron spectrum (not shown in Table~\ref{tbl:collective-electronic-excitations}) is the partial contribution of double excitations in the B$\rightarrow$B transitions.
Such excitations are also partially present in the 3-rd excited state of \twot{} but somehow disappear in \twoo{}.
The "pure" double electronic transitions in all the investigated systems are presented in the upper part of the spectrum, as shown in Table S9 of the ESI\dag.
\\
\textbf{Analysis of the leading contributions. }
For \onet{}, all three lowest excitations have leading contributions from the $\pi_{\rm B}\rightarrow\pi_{\rm B}^*$, where B indicates the benzenoid rings. 
They differ between themselves in the admixture of transitions from the nitrogen lone pair (LP$_{\rm N}$) orbital to the $\pi_{\rm B}^*$ and $\sigma^*$ orbitals. 
In the second and third excited state of \onet{}, transitions of the $\pi_{\rm B}\rightarrow\sigma^*$ character appear additionally.
Upon polymer elongation (\oneo{}), the excitations become almost solely dominated by the $\pi_{\rm B}\rightarrow\pi_{\rm B}^*$ transitions.

The electronic spectrum of \twot{} is very complex and involves transitions of many characters.
The leading contributions for the first excited state come from the nitrogen lone-pairs (LP$_{\rm N}$), $\pi_{\rm N}$, and $\pi_{\rm B}$ orbitals to the $\pi^*_{\rm N}$ orbital (where the subscript N underlines that the orbital is centered at the nitrogen atom). 
Additionally, we find smaller but non-negligible contributions of type LP$_{\rm N}\rightarrow\pi^*_{\rm N}$, $\pi_{\rm B}\rightarrow\pi^*_{\rm N}$, and $\pi_{\rm N}\rightarrow$ $\pi^*_{\rm N}$, where the index Q indicates the quinoid ring.
LP$_{\rm N}\rightarrow\pi_N^*$ and $\pi_{\rm B}\rightarrow\pi^*_{\rm N}$ electronic transitions dominate the second excited state and $\pi_{\rm B}\rightarrow\pi^*_{\rm B}$ the third excited state of \twot{}, respectively.
Moving to \twoo{}, we observe a more organized spectrum composed of less diverse transitions.
Specifically, these are $\pi_{\rm Q}\rightarrow\pi^*_{\rm Q}\backslash\pi^*_{\rm N}$ and $\pi_{\rm B}\rightarrow\pi^*_{\rm Q}\backslash\pi^*_{\rm N}$ transitions for the first,  LP$_{\rm N}\rightarrow\pi^*_{\rm N}$, $\pi_{\rm Q}\rightarrow\pi^*_{\rm N}$, and $\pi_{\rm B}\rightarrow\pi^*_{\rm Q}$ transitions for the second, and the LP$_{\rm N}\rightarrow\pi^*_{\rm N}$ and $\pi_{\rm B}\rightarrow\pi^*_{\rm N}$ transitions for the third excited state, respectively.
Thus, the elongation of emeraldine (\textbf{2}) profoundly affects its low-lying electronic transitions, revealing the involvement of quinoid rings only in the octamer configuration.

The electronic spectrum of \threet{} is as complex as \twot{}, differing mainly in the increased involvement of quinoid orbitals and the presence of $\sigma_N$ orbitals (see the corresponding orbitals in Figures S10-S18 of the ESI\dag).
The first excited state of \threet{} is dominated by LP$_{\rm N}$$\rightarrow\pi^*_{\rm Q}$, $\sigma_{\rm Q}\rightarrow\pi^*_{\rm Q}\backslash\pi^*_{\rm N}$, and $\pi_{\rm Q}\rightarrow\pi^*_{\rm Q}$ transitions, the second one by LP$_{\rm N}\rightarrow\pi^*_{\rm N}$, $\sigma_{\rm Q}\rightarrow\pi^*_{\rm Q}$, and $\pi_{\rm B}\rightarrow\pi^*_{\rm Q}$ transitions, and the third one by LP$_{\rm N}\rightarrow\pi^*_{\rm Q}$, $\sigma_{\rm Q}\rightarrow\pi^*_{\rm Q}$, and $\sigma_{\rm N}\rightarrow\pi^*_{\rm N }\backslash\pi^*_{\rm Q}$ transitions.
The electronic spectrum of the corresponding structure \threeo{} is less complex, dominated by three main types of transitions.
Specifically, these are $\pi_{\rm Q}\rightarrow\pi^*_{\rm N}$ and LP$_{\rm N}\rightarrow\pi^*_{\rm N}$ transitions for the first excited state, LP$_{\rm N}\rightarrow\pi^*_{\rm N}$, $\pi_{\rm Q}\rightarrow\pi^*_{\rm N}$, and $\pi_{\rm B}\rightarrow\pi^*_{\rm N}$ transitions for the second one, and LP$_{\rm N}\rightarrow\pi^*_{\rm N}$ and $\pi_{\rm B}\rightarrow\pi^*_{\rm N}$ transitions for the third one, respectively. 
\subsubsection{Orbital-pair correlation analysis}
\begin{figure*}
    \centering
    \includegraphics[scale=1.3]{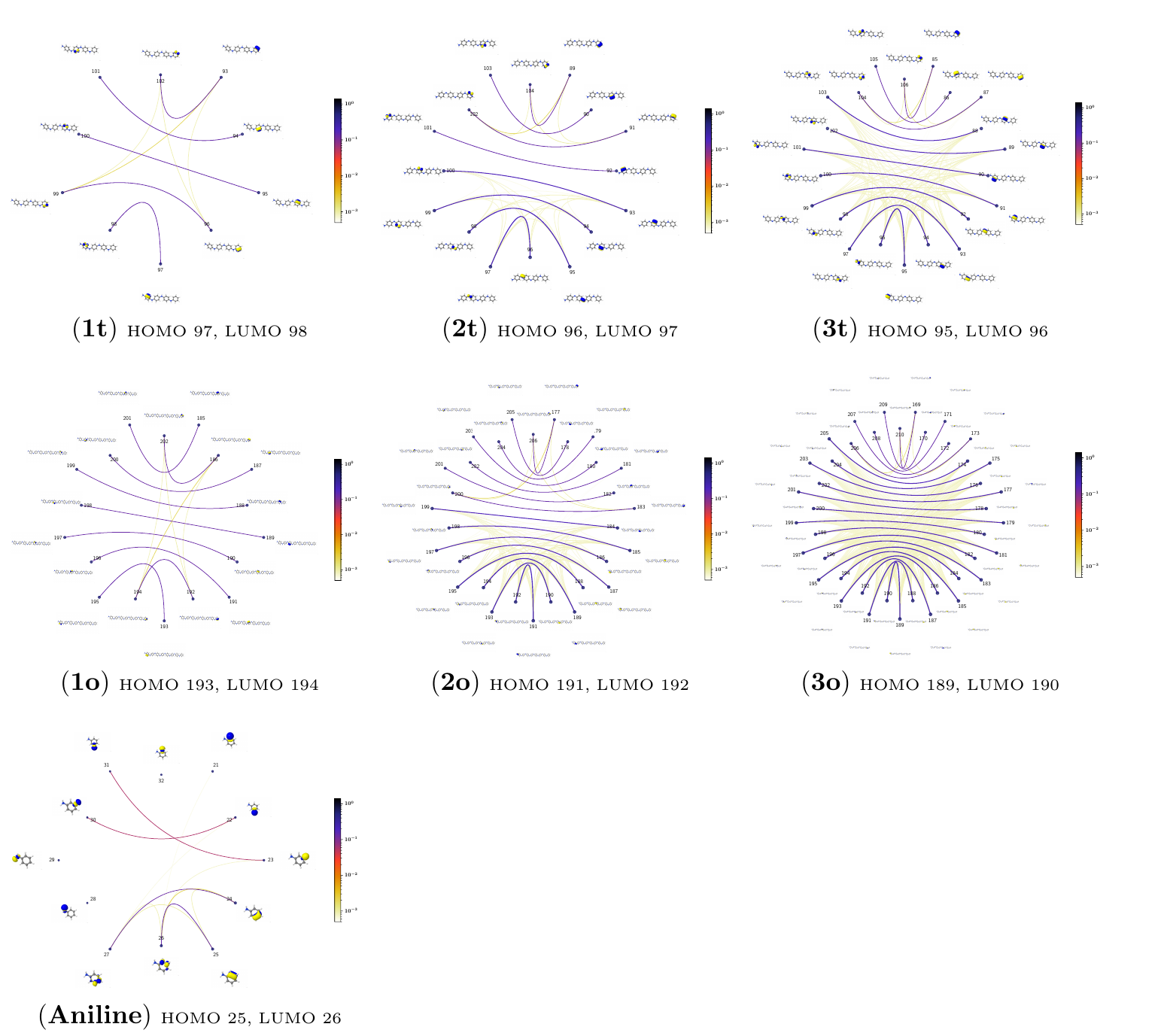}
        {\tiny}
    \caption{The orbital-pair mutual information for aniline and all investigated PANIs in their tetramer ({\bf{t}}) and octamer ({\bf{o}}) forms calculated from the pCCD ground-state wavefunctions within the \textsc{PyBEST} software package.}
    \label{fig:mutual-information}
\end{figure*}

To better understand the electronic structures and the structure-to-property relationship of the investigated PANIs, we performed an orbital-pair mutual information analysis depicted in Figure~\ref{fig:mutual-information}.
The strength of the mutual information (that is, orbital-pair correlations) is color-coded in Figure~\ref{fig:mutual-information}. 
Furthermore, only the most strongly-correlated orbital pairs are shown for better visibility.
To a large extent, these are classical correlation effects.~\cite{qit-concepts-schilling-jctc-2021}
Interestingly, all the investigated systems have the most correlated orbitals around the valence region (the benzenoid/quinoid ring).  
These are the $\pi$ and $\pi^*$ orbital combinations, including the HOMO--LUMO pairs. 
They do not coincide entirely with the pCCD orbitals involved in the electronic excitations. 
The pCCD orbitals are optimal for the ground but not necessarily for excited state structures. 
For aniline, we observe only two strongly correlated pairs, HOMO--LUMO and HOMO-1--LUMO+1. 
For \onet{}, we have five pairs, for \oneo{} nine pairs, for \twot{} eight pairs, for \twoo{} 15 pairs, for \threet{} eleven pairs, and for \threeo{} 21 pairs. 
The stronger $\pi$-$\pi^*$ orbital pairs are present in the oxidized forms of PANIs and longer polymer chains. 
That is a clear indication of increased conjugated properties in such systems and correlates with the analysis of low-lying 
part of their electronic spectrum. 

The quantum information analysis of PANI structures points to an increased multi-reference character in longer polymer chains.
That is highlighted by the growing number of strongly-correlated orbitals in Figure~\ref{fig:mutual-information}.
Thus, we anticipate that for such structures, pCCD-based methods should be superior to DFAs.

\section{Conclusions}\label{sec:conclusions}
In this article, we employed modern quantum chemistry methods to investigate the electronic structures and properties, such as vibrational and electronic spectra, of the aniline molecule and PANIs at different oxidation states and lengths. 
We analyzed their structure-to-property relationship for the first time. 

The BP86-optimized electronic structures and vibrational frequencies of aniline and PANIs are in excellent agreement with the available experimental data indicating the right choice of the xc functional. 
The characteristic structural and vibrational features of PANIs in the tetramer form (\onet, \twot{}, and \threet) are almost indistinguishable from their octamer counterparts (\oneo, \twoo~and \threeo).
Thus, the tetramer forms of PANIs are adequate models for longer polymer chains when considering structural and vibrational features, regardless of their oxidation states. 
However, the length of the PANI chain profoundly affects the electronic spectra and the overall electronic structure. 
Moving from aniline to polymeric structures, the mutual information analysis indicates the increased multi-reference character of the systems.
Such observation calls into question the reliability of DFAs in predicting ground and excited-state properties in a balanced way.
A numerical indication is already observed for the CAM-B3LYP xc functional having convergence issues for octamer structures (\oneo, \twoo{}, and \threeo).
As an alternative, we propose to use pCCD-based methods that utilize the complete set of variationally optimized orbitals at the correlated level and can cope with such complex electronic structures.  
An additional advantage of pCCD-based methods is the optimization of all orbitals on an equal footing (up to a thousand basis functions in this work).
The final pCCD orbitals are localized.
The excitation energies are composed of many small components for such localized orbitals.
We showed that working with localized orbitals (like pCCD-optimized ones) allows us to dissect the collective CT and L character of electronic transitions in each PANI for the first time. 
Specifically, we demonstrated that EOM-pCCD+S electronic spectra of emeraldine and pernigraniline have a dominant CT character and that polymer elongation changes the character of the leading transitions. Such an analysis is not possible using the delocalized canonical DFT orbitals.
Our results highlight the strong structure-to-property relationship for electronic excitations, where the character of the excited states changes upon polymer elongation of the oxidized forms of PANIs.
For instance, elongating the polymer \textbf{2} delocalizes the leading transitions of the first excited state over the whole quinoid ring (\textbf{2o}), while \textbf{2t} features leading transitions to the quinoid $\pi_{\rm N}^*$ orbital.
Similarly, the first excited state in \textbf{3} changes its character upon polymer elongation.
While \textbf{3t} features more delocalized leading transitions from the LP$_{\rm N}$ to the quinoid rings, the dominant transitions in \textbf{3o} are centered on the quinoid ring. 

Finally, our work underlines the potential of pCCD-based methods in modeling organic electronics and motivates their further development.
Based on previous studies,~\cite{eom-pccd,eom-pccd-lccsd} EOM-pCCD-based models provide reliable excited states' characters, while the overall excitation energies might be too high compared to experimental results.
We should stress that adding dynamical correlation did not significantly improve excitation energies in polyenes.~\cite{eom-pccd-lccsd}
This observation suggests that other effects like basis set size or environmental effects have to be considered. One possibility to account for environmental effects is to use embedding methods.~\cite{pccd-embedding}
Such investigations are currently in progress in our laboratory.
\section{Conflicts of interest}
There are no conflicts to declare.
\section{Acknowledgment}\label{sec:acknowledgement}
S.J. and P.T.~acknowledge financial support from the SONATA BIS research grant from the National Science Centre, Poland (Grant No. 2021/42/E/ST4/00302). P.T.~acknowledges the scholarship for outstanding young scientists from the Ministry of Science and Higher Education.
The research leading to these results has received funding from the Norway Grants 2014--2021 via the National Centre for Research and Development.

The authors thank Julia Romanova for providing us with the initial xyz structures of polyanilines. 
Calculations have been carried out using resources provided by the Wroclaw Centre for Networking and Supercomputing (http://wcss.pl), grant no. 411. 
\providecommand*{\mcitethebibliography}{\thebibliography}

\end{document}